# Natural networks

Tuomo Hartonen[a,d] and Arto Annila[a,b,c]

[a]Department of Physics, [b]Institute of Biotechnology and [c]Department of Biosciences, FI-00014 University of Helsinki, Finland;
[d]Department of Applied Physics, Aalto University School of Science and Technology, FI-00076 AALTO, Finland

ABSTRACT
Scale-free and non-computable characteristics of natural networks are found to result from the least-time dispersal of energy. To consider a network as a thermodynamic system is motivated since ultimately everything that exists can be expressed in terms of energy. According to the variational principle, the network will grow and restructure when flows of energy diminish energy differences between nodes as well as relative to nodes in surrounding systems. The natural process will yield scale-free characteristics because the nodes that contribute to the least-time consumption of free energy preferably attach to each other. Network evolution is a path-dependent and non-deterministic process when there are two or more paths to consume a common source of energy. Although evolutionary courses of these non-Hamiltonian systems cannot be predicted, many mathematical functions, models and measures that characterize networks can be recognized as appropriate approximations of the thermodynamic equation of motion that has been derived from statistical physics of open systems.

## 1. Introduction

On-going global integration gives rise to numerous networks, most noticeably in telecommunication and transportation that, in turn, arrange social networks and structure distribution of work. It is a striking observation that networks are all alike in their principal properties (1). Scale-free characteristics are ubiquitous. Not only the infrastructure of socio-economic systems (2,3,4) but also biological networks, e.g., metabolic (5), gene and protein regulatory networks (6) as well as cognitive (7) and population networks (8) display power laws (9). Likewise, power laws dominate degree distribution of interaction networks of physical systems (10) that range from Bose-Einstein condensates (11) to percolation of galaxies (12).

Universality implies that network proportionate progression by preferential attachment (13,2) is a natural process, i.e., a manifestation of the supreme law of nature. This profound principle is known by many names, best as the principle of least action (14) and the second law of thermodynamics (15). Also the maximum power principle (16), Yule's process for cumulative advantage (17) and evolution by natural selection (18) can be recognized as accounts of the probable process for the least-time dispersal of energy (19,20).

Considering the irrefutable imperative in energy transduction, it becomes apparent why numerous natural networks also display ubiquitous scale-free and non-deterministic characteristics of least-time free energy consumption. However, our objective in this study is not to propose a new mathematical model to account for the evolution of natural networks. Rather, we wish to promote an old, holistic, physical portrayal of nature to clarify why certain mathematical functions, distributions and models, as well as measures are so successful in modeling and characterizing networks. We will not question the natural law itself, but only analyze its equation of motion to draw conclusions about the universality of network qualities. In this way, we hope to communicate why prominent patterns propagate throughout nature.

## 2. Natural processes

The essence of physics is to subsume specific details of distinct systems into universal principles. To this end, the principle of least action in its original holistic and hierarchal form (14) describes a system within surrounding systems in least-time progression toward a free energy minimum. Differences in energy will level off as soon as possible when flows of energy search and naturally select to direct from highs to lows along the paths of highest throughputs, known as geodesics (18,20). The irrevocable least-time consumption of free energy results in sigmoid courses of growth or decline as well as skewed, nearly log-normal distributions (21,22,23,24). Also oscillatory, chaotic and non-deterministic behavior (25,26,27) as well as power-law scaling and branching are qualities of natural systems (28,29,30,31,32,33) that emerge from the universal quest for least-time energy dispersal (34,35).

The notion of a network is a powerful portrayal of an energy transduction system. Nodes represent repositories of potential energy and links correspond to paths for flows of energy. We find this association between a physical network and its graphical representation motivated, since all systems must embody energy, at least one quantum, to exist. Moreover, the physical picture of a network complies with conservation of energy and causality. A flow of energy along a link stems from one node upstream that has opened itself up and expelled at least one quantum. The quantum will eventually be captured by another node downstream as it closes to a new stationary-state action. Thus, from a physical perspective, natural networks will emerge and evolve in the quest for least-time consumption of scalar and vector potential differences, i.e., components of force. An evolving network will prefer attachments that will further the most effective free energy consumption. A particular network topology results from the natural process in particular circumstances.

The scale-independent and non-computable characteristics of natural networks will be unraveled when the equation of motion for the least-time energy dispersal is formulated and analyzed. In the context of network theory, each node is regarded as a quantized repository of energy. It is indexed with $j$ and assigned with energy density $\phi_j = N_j \exp(G_j/k_BT)$, where $N_j$ denotes the number of constituents (quanta) associated with scalar potential $G_j$ relative to the average energy density $k_BT$ per node of the network system (36). Since the components $N_j$ of the $j$-node are explicitly denoted, the formalism is self-similar. Accordingly, any constituent at a lower level of hierarchy can also be regarded as a node (Fig. 1).

The probable process diminishes energy differences between the nodes and their respective surroundings. It is driven by the consumption of free energy terms $A_{jk} = \Delta\mu_{jk} - i\Delta Q_{jk}$. The two components of $A_{jk}$ comprise the mutual differences in scalar potentials, known also as chemical potentials, $\Delta\mu_{jk} = k_BT(\ln\phi_j - \ln\phi_k)$, as well as differences in vector potentials, more commonly referred to as dissipation $\Delta Q_{jk}$. The imaginary unit merely emphasizes that the scalar and vector potentials are orthogonal to each other (37,38,39). Moreover, it is noteworthy that when the state of a node changes, at least one quantum will either be absorbed or emitted. Only reversible exchange of quanta will leave a pair of nodes intact, i.e., stationary. The flows of energy between the nodes are literally inter-actions, since each node is regarded as a physical system characterized by its action and associated symmetry (40,41).

The energetic status of a network system can be derived from statistical mechanics of open systems (20,35,37). The additive hence logarithmic probability $P$ measure of the network, known as entropy

$$S = k_B \ln P = k_B \sum_j \ln P_j \approx k_B \sum_j N_j \left(1 - \sum_k A_{jk}/k_BT\right) \tag{1}$$

contains the bound $k_BT\Sigma N_j$ and free $\Sigma N_j A_{jk}$ forms of energy. The Stirling's approximation for indistinguishable combinations $\ln N_j! \approx N_j(\ln N_j - 1)$ implies that $k_BT$ is a sufficient statistic for distribution of energy, i.e., $A_{jk}/k_BT \ll 1$. In other words, the heat capacity of the network is big enough so that absorption or emission of one quantum will not cause a marked change in the average energy density $k_BT$. The statistical approximation will hold for the entire network, but not for a specific node when it is about to bud or go extinct, or when new links branch out (32,42). Then, the critical step is best characterized directly by the change $dP_j/dt$. Moreover, the change of $P$ will be abrupt at a phase transition when the system rapidly reorganizes its entire interaction network (43).

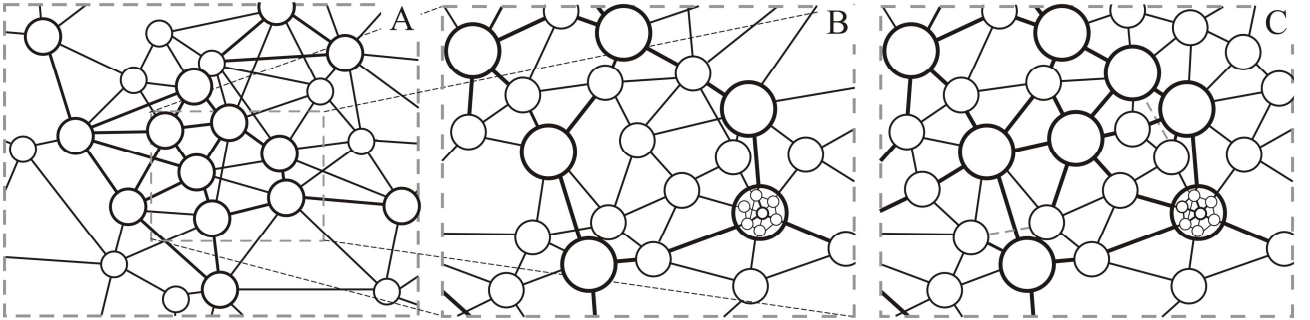

**Fig. 1.** (A) Natural network is a self-similar transduction system of pathways (links) and repositories of energy (nodes). (B) The expansion exemplifies that the links comprise nodes too. Also each node is a thermodynamic system of its own internal interactions between its constituent nodes. (C) The network will evolve when nodes are subject to differences of energy. Ensuing net flows of energy will cause changes in the network, e.g., by altering transduction capacity of links as well as attaching new nodes or eventually discarding old less effective connections.

According to the variational principle, flows of energy will themselves search for paths and eventually also open new links to consume free energy in the least time (20,44). Hence the network notion of preferential attachment is subsumed in the natural selection for the least-time dispersal of energy. Consequently, entropy will not only increase,

$$\frac{dS}{dt} = k_B L \geq 0 \ ; \quad L = -\sum_{j,k} \frac{dN_j}{dt} \frac{A_{jk}}{k_B T} \quad (2)$$

but it will increase in the least time. It is of interest to note that the least-time imperative is equivalent to Newton's second law $\mathbf{F} = d\mathbf{p}/dt$, which says that the change in momentum $\mathbf{p}$ will keep directing along the resultant force $\mathbf{F} = \Sigma \mathbf{F}_j$ (14,44). Thus, at all times and at all places, the network will naturally select pathways to evolve so that free energy cannot be consumed any faster.

The conservation of energy requires that an influx ($\Sigma A_{jk} < 0$) or efflux ($\Sigma A_{jk} > 0$) will force the node to change its constituents at a rate (20)

$$\frac{dN_j}{dt} = -\sum_k \sigma_{jk} \frac{A_{jk}}{k_B T} \quad (3)$$

proportional to free energy. The coefficient of conductance $\sigma_{jk}$ is a characteristic of the link between the nodes indexed with $j$ and $k$. For example, a city will grow due to an influx of inhabitants from surrounding rural areas. According to self-similar formalism, the link itself can be considered as a network of nodes and links (Fig. 1). For example, when computing is distributed, a client and a server do not usually link directly, but instead they link over a network of hubs and connections. Likewise, two cities are rarely linked by a non-stop train connection, as the train also stops at other major towns for influx and efflux.

According to the variational principle, energy tends to flow along least-time links. The least action defines the length $s$ of geodesic in energetic terms by $2K = \int (ds/dt)^2 dt$ (45) equivalent to

$$\left(\frac{ds}{dt}\right)^2 = \frac{d2K}{dt} = T\frac{dS}{dt} = -\sum_{j,k} \frac{dN_j}{dt} A_{jk} = \frac{1}{k_B T} \sum_{j,k} \sigma_{jk} A_{jk}^2 \geq 0 \ . \quad (4)$$

as proportional to the magnitudes of free energy components. Accordingly, when there is no difference in energy between the $j$- and $k$-node, the two nodes will be indistinguishable from each other, hence $j = k$ and the particular $s_{jj}$ vanishes. The average of $s = \Sigma s_{jk}$, as a characteristic of network topology, will decrease with the increasing number of $jk$-links between an invariant set of nodes. It is noteworthy that a particular value of $s$ cannot be determined for an evolving network because the total energy of the system is changing. Physically speaking, eigenvalues and eigenmodes cannot be determined when they are changing.

If transduction rates $dN_j/dt$ were suboptimal, counterforces $A_{jk}$ would rise and redirect the flows of energy back along the most voluminous gradients. In other words, even when a sufficiently statistical network is evolving, its distribution of energy is not expected to depart much from a quasi-stationary balance known as Le Chatelier's condition (46)

$$A_j = \sum_k A_{jk} \approx 0 \Leftrightarrow N_j \approx \prod_k \left(N_k e^{-(\Delta G_{jk} - i\Delta Q_{jk})/k_B T}\right) \quad (5)$$

where $\Pi_k$ is over all $k$-substrates. The product form in Eq. 5 reveals that the $j$-node materializes from $k$-multiplicative operations. The multiplicative form is characteristic of a log-normal distribution (47) whose cumulative curve follows a power law. The distribution's dependence on the average energy is familiar from the temperature dependence of the Maxwell-Boltzmann velocity distribution and from the black-body radiation spectrum, but it is also recognized in temporal changes during ecological succession (48,49), economic development (50,51), cultural changes (52,53) as well as when logistic (54) and communication infrastructure are building up (55).

The above thermodynamic description of networks by the natural law is formally simple, yet its analysis (Eq. 2) reveals that evolution of a network is an intractable process. Namely, when a particular source of energy, i.e., a node is consumed via two or more links, i.e., degrees of freedom, the flows of energy and the energy difference cannot be separated from each other to solve Eq. 2 by way of integration to a closed form (20,44).

It is a mere consequence of conservation of energy that an evolutionary step as a dissipative event, just as a developmental step, will alter both the system and its surroundings. Due to this intrinsic interdependence among all constituents of the system and its surroundings, evolution changes its settings, i.e., the energy landscape that directs the natural process. The flow itself will affect conduction by urging an increase in communication capacity or by strengthening communication lines, such as synapses of neurons. Likewise, a river itself will erode the landscape by the mere act of flowing, and thus affect its own flow.

Due to dissipation, the change in momentum is not collinear with the velocity, which is a characteristic of non-Abelian systems. In other words, natural processes are dissipative, path-dependent and inherently intractable. It is noteworthy that the non-deterministic nature of network evolution does not stem from network complexity as such, but appears already in the problem of three bodies (56) and other hard problems with two or more degrees of freedom (57,58,59). Although evolution in general is a non-computable process, certain mathematical models of networks can be solved (60,61).

Finally, the above conclusions do not depend on how one defines a network system. When a definition of a network happens to include nodes that are in imbalance with each other, the development will manifest primarily as a restructuring of the network when internal forces are being consumed. For example, social systems display this sequence of events during integration processes of immigrants. Likewise, the conservation of energy in bound and free forms of interactions will be respected when a definition of a node happens to subsume nodes. For example, a definition for two cities may seem arbitrary in subsuming some suburban communities while discarding others. Nevertheless, fervent communication between the twin cities will establish a metropolitan area irrespective of its formal boundaries.

### 3. Approximate forms of natural degree distribution

The above thermodynamic formulation can be analyzed to reveal the ubiquitous characteristics of natural networks. In particular the skewed degree distribution can be found as an excellent approximation of the thermodynamic stationary-state condition ($d\ln P = 0$) of Eq. 5.

$$\ln N_j = \ln \prod_k \left( N_k e^{-(\Delta G_{jk} - i\Delta Q_{jk})/k_B T} \right) = j \ln N_1 \sum_{1 \leq m,n \leq j} -A_{mn}/k_B T \propto j \ln N_1, \tag{6}$$

which is linear on a semi-log scale (24,34). Here any $j$-node in the hierarchy of the network (Fig. 1) is expressed as being composed of some basic constituents $N_1$ (quanta), because all nodes are results of some earlier processes. Then it follows from this recursive form that the $j$-node with $N_j$ constituents embodies an energy density

$$\phi_j = N_j e^{G_j/k_B T} = N_1^j e^{j(G_1 + i\Delta Q_1)/k_B T} = e^{j(\ln \phi_1 + i\Delta Q_1/k_B T)} \Leftrightarrow \ln \phi_j = j \ln \phi_1', \tag{7}$$

where the number of quanta $j\Delta Q_1$ that have been incorporated in the assembly of $\phi_j$ are included in the shorthand notation $\phi_1'$. Accordingly, another node with $j+n$ constituents comprises an adjacent energy density

$$\phi_{j+n} = \exp\left[(j+n)\phi_1'\right] = \phi_j \exp\left(n \ln \phi_1'\right) \Leftrightarrow \ln \phi_{j+n} = (j+n) \ln \phi_1'. \tag{8}$$

This form reveals that when $n \ll j$ the distribution of energy densities $\phi_{j-n\ldots j+n}$ over a range of nodes $j-n \ldots j+n$ about $\phi_j$

$$\ln \phi_{j-n\ldots j+n} = \ln \phi_j + \sum_n n \ln \phi_1' \tag{9}$$

is normal according to the central limit theorem. The condition of small variation is effectively the criterion by which the nodes are qualified to the same degree distribution. For example, when a distribution of cities is compiled from a network of population centers, small villages or suburbs will be excluded. Due to the ubiquitous quest for least-time free energy consumption, natural distributions will display scale-free, skewed characteristics irrespective of classification criterion.

It is noteworthy that the natural distribution does deviate from the aforementioned log-normality in the way it tails off both at low and high ends. The distribution will tail off when the functional mechanism of a particular class of nodes becomes increasingly ineffective or energetically expensive as a means of energy transformation. For example, powerful energy transduction mechanisms that are characteristics of a city such as factories are not supported by a small village. Conversely, an increasingly larger metropolis struggles with increasingly more acute transportation problems that curtail its further growth. Due to these imperatives in energy transduction, the sigmoid cumulative curve will deviate from the power law at both the low and high ends. The low end cut-off is usually referred to as the finite-size effect (62). For example, the species-area relationship (21,63,64,65,66) is a well-known cumulative curve of ecosystems which totals from distributions of species that populate increasing larger areas (67). It mostly follows a power law, but at the low end the resources in a small area do not support any species in a particular genus, as well as toward the high end no species of that genus is capable of harvesting the large but scattered resources (68,69,70,71,72,73,74).

The recursive power-law form of the cumulative probability distribution is closely followed by certain functions. For example, beta distribution for large values of either of its gamma-function arguments is a good approximation of certain natural distributions (75). Moreover, the scale-free stationary distribution $\Sigma k^{-\alpha}$ is proportional to the Riemann zeta function $\zeta(\alpha)$, which in turn has been associated to the stationary states by the thermodynamic principle (76). However, these and other mathematical models (17) rarely account for the entire span of a natural distribution that tails off at both ends due to the mechanistic limitations of energy transduction.

## 4. Kinetic models of evolving networks

Evolution of natural networks mostly follows a power law. This time-dependence can be recognized to result from maximal energy dispersal when the equation for evolving probability $dP/dt = LP$ (Eq. 2) is analyzed.

In practice, it is the number of nodes $N_j$ that can be monitored during evolution rather than the associated probability $P_j$ (Eq. 2). When $N_j$ of the nascent network is small in comparison with $N_j^{ss}$ of the mature network, the change in the free energy $\Sigma_k \partial A_{jk}/\partial t = \Sigma_k \sigma_{jk}$ is a good approximation independent of the energy flow. This deterministic, zero-order approximation

$$\sum_k \frac{\partial A_{jk}/k_B T}{\partial t} = \frac{dN_j}{dt} \sum_k \frac{\partial A_{jk}/k_B T}{\partial N_j} \approx \sum_k \sigma_{jk} \Rightarrow \frac{dN_j}{dt} = \sigma_j N_j, \text{ when } A_{jk}(t) \approx A_{jk}(0) \quad (10)$$

will give the exponentially increasing initial growth $N_j(t) = N_j(0)\exp(\sigma_j t)$ when variables are separated and integrated from 0 to $t$. Evolution will punctuate off when a transformation mechanism appears in the system for the first time and taps into a nascent reservoir of free energy (21). For example, the initial growth of a business branch is exponential.

Conversely, the decreasing exponential approximation will be obtained when energy in the maturing population $N_j(t)$ has almost attained $N_j^{ss}(\infty)$ at the stasis where $A_{jk}^{ss} = 0$. Then, the change $\Sigma_k \partial A_{jk}/\partial t = -\Sigma_k \sigma_{jk}$ is nearly constant

$$\sum_k \frac{\partial A_{jk}/k_B T}{\partial t} = \frac{dN_j}{dt} \sum_k \frac{\partial A_{jk}/k_B T}{\partial N_j} \approx -\sum_k \sigma_{jk} \Rightarrow \frac{dN_j}{dt} = -\sigma_j N_j, \text{ when } A_{jk}(t) \approx A_{jk}^{ss} = 0. \quad (11)$$

The exponential decrease $N_j(t) = N_j^{ss} - N_j(0)\exp(-\sigma_j t)]$ at the late stage will be obtained when variables are separated and $N_j$ is integrated from $N_j(0)$ to $N_j^{ss}$. Evolution will settle to a stasis when the transformation mechanisms have consumed all free energy. For example, a mature business branch will saturate a market. Thereafter, the operation depends only on the amount of steadily available and renewable potential.

In the intermediate region between the initial increase and the final decrease, the population $N_j$ is given by Eq. 6, which is valid for a sufficient statistical system. This means that $N_j$ of the quasi-stationary network can be written in terms of multiplicative operations of the basic constituents in numbers $N_1$,

$$N_j = \prod_k \left(N_k e^{-A_{jk}/k_B T}\right) = N_1^j \prod_{1 \leq m,n \leq j} e^{-A_{mn}/k_B T} = \alpha_j N_1^j, \quad (12)$$

where the constant $\alpha_j = \Pi_{m,n}\exp(-A_{jk}/k_B T)$ is over available $m,n$-indexed transformation paths. The multiplicative form is recognized as a power law. The typical time course about a quasi-stationary point $A_{jk}^{qs}$ follows from the approximation $\Sigma_k \partial A_{jk}/\partial t = \partial A_j/\partial t = -\sigma_j$

$$\frac{dN_j}{dt} = \frac{dN_j}{dN_1}\frac{dN_1}{dt} = \alpha_j j N_1^{j-1}\frac{dN_1}{dt} = \frac{jN_j}{N_1}\frac{dN_1}{dt} \Rightarrow \frac{dN_j}{N_j} = \frac{jdN_1}{N_1}, \text{ when } A_{mn} \approx A_{jk}^{qs}. \qquad (13)$$

When the variables are separated, the integration will give $\ln N_j = j\ln N_1 +$ a constant. The characteristic of scale invariance is apparent from the log-log plot where the curve is a straight line.

The obtained functional form is also familiar from the law of mass-action. However, it is noteworthy that it is not the number of nodes but the energy contained in the nodes that contributes to the driving force of evolution $A_{jk}/k_BT$. Consequently, the law of mass-action does not comply with conservation of energy. Hence, in that model of network kinetics (77) the forward and backward flow coefficients are erroneously deemed as if they were changing during the course of evolution, whereas in reality it is free energy that is decreasing and the conduction coefficient $\sigma_{jk}$ in Eq. 3 is a constant. Obviously, new means of transformation may also emerge to facilitate the flows.

Finally, we emphasize that the evolutionary equation Eq. 2 is non-deterministic, unlike its mathematical models that can be integrated to closed forms. Often the overall sigmoid course is, to a good approximation, also given by the logistic equation. The above analysis of the evolutionary equation (Eq. 2) and the associated kinetic (Eq. 3) and balance equations (Eq. 5) for the initial, intermediate and final stages of growth (or decline), reveals that the ubiquitous power laws are consequences of the natural principle of the least-time energy dispersal.

## 5. Growth models of natural networks

The preferential attachment model will generate random scale-free networks with skewed degree distributions (2,78,79). The basic algorithm weights the connection probability $P_j(N)$ of a $j$-node with its degree ($N$). Hence, new nodes are most likely to make connections with already densely connected nodes. This method mimics the natural process of least-time energy dispersal where interactions among constituents of a larger and denser system form increasingly more effective mechanisms of energy transduction. The basic algorithm will reproduce a power-law region, but not the early punctuation and the late settling to stasis, since the thermodynamic limitations of growth denoted in Eqs. 1 and 2 are not included in $P_j$. After all, it is not the number of links that drives the growth but the energy influx via links that fuels the expansion.

Weighted networks (80) are better models to account for variation in energy transfer characteristics of natural links. When the $j$-node is assigned with strength

$$s_j = \sum_k w_{jk} \qquad (14)$$

as a weighted sum of links to neighboring $k$-nodes, the degree distribution $P(N)$ is replaced by a strength distribution $P(s)$. This model can be recognized as an approximation of Eq. 3 when assuming that the driving forces of evolution $A_{jk}$ would be constant and small relative to the overall energy content, i.e., $A_{jk}/k_BT \ll 1$. Then $s_j$ would be equal to the total conductivity $\sigma_j = \Sigma\sigma_{jk}$ that links the $j$-node with its surrounding $k$-nodes.

When the network algorithm assigns the $j$-node with probability

$$P_j = s_j \Big/ \sum_i s_i \qquad (15)$$

to attach to a new $k$-node (81), the basic idea of the preferential attachment that "rich get richer" will be transcribed for the weighted networks in a form that "busy get busier". For example, when a new highway is constructed between two cities, traffic on older and smaller roads will reduce. Simple weight-driven dynamics will yield scale-free characteristics (82). Since an added link introduces variations to the existing weights across the network, the algorithm conforms to the interdependence among natural nodes, although Eq. 15 does not express $P_j$ in explicit terms of energy as Eq. 1 does. Normalization by the sum of weights relates to the average energy per node. When $P_j$ is bound, the distribution will be stationary. However, since the net influx powers the growth of the natural network, $k_BT$ will increase during evolution, hence there is no firm ground for normalization. When energy is absorbed into a network, approximately logarithmic progression of both the degree and the weight of a node $i$ (81) will follow. For example, when a city becomes more prosperous, more and more people will move in and the countryside will become desolate. The urbanization will also result to increase in traffic in and out of the city thus strengthening the important connections between other growing areas. However, growth of degree and weight always requires influx of energy, i.e., insertion of new nodes.

**6. Topological measures of natural networks**

In thermodynamic terms, the lengths of geodesics (Eq. 4) are informative about the network topology, whereas in network theory the average length

$$l = \frac{1}{N(N-1)} \sum_{j,k} d_{jk} \qquad (16)$$

of a path in a non-weighed graph is defined as the sum of the shortest distances $d_{jk}$ between all combinations of $j$- and $k$-nodes in total $N(N-1)$. This measure parallels the total geodesic length when normalizing with all conceivable $jk$-combinations. However, according to the thermodynamic tenet the flows of energy themselves do value a link by its means $\sigma_{jk}$ and associated driving forces $A_{jk}$. For example, when a logistic network is structuring itself across a rough terrain, means of transportation and expected returns will matter more when deciding which lines of transportation require upgrading than the actual distances as the crow flies. When weights are applied on the links, reality will be modeled more precisely.

Clustering coefficients are informative about local linkage density and eventually, when reduced to an average figure of merit, also about the average connectivity of the entire network (83). In thermodynamic terms, the skewed degree distribution as a quasi-stationary partition (Eq. 5) covers the network of diverse localities. However, this measure is not normalized with all conceivable connections. In general, the statistical mechanics of open systems refrain from normalization because when the total energy content of an evolving system is changing, and even in a non-deterministic manner, there are no grounds for normalization.

Finally, the logarithmic dependence $L \propto \ln N/\ln k$ of the typical distance $L$ between two randomly chosen nodes on the total number of nodes $N$ and the number of neighbors $k$ also deserves to be related to thermodynamic terms. When a locus is understood as a closed action of scalar potential energy, a distance between two loci will also be understood, e.g., as a difference between one chemical potential $\mu_j \propto \ln N_j$ and the other $\mu_k$. Thus, the typical distance is proportional to the potential of an entire network $\mu \propto \ln N$. Moreover, the typical distance scales down with the logarithm of the number of neighbors in a small-world network, because the nearby high-throughput links hardly contribute to the length of a geodesic (Eq. 5). Thus, the total number of nodes is effectively scaled down by the number of neighbors.

Obviously, the numerical values of topological measures will change when the network is evolving, but the thermodynamic tenet emphasizes that evolution is by its nature non-deterministic. This is of course understood in practice, where trends are followed and extrapolated with reservations rather than attempting to make precise predictions.

**7. Conclusions**

The physical portrayal of networks as energy transduction systems may appear to some superficial by subsuming numerous mechanistic details in its general concepts, but the ubiquitous scale-free and non-deterministic characteristics themselves imply that there is an underlying universal law in action. The principle of least action, when given as an equation of motion and analyzed, will reveal that network structure and evolution can indeed be outlined by certain analytical functions, notably by skewed, log-normal distributions, power laws and logistic equations. However, these functional forms are reticent in revealing the underlying cause of universality, i.e., that energy differences drive natural networks toward free energy minima in respective surroundings and that the ubiquitous characteristics reflect the least-time dispersal of energy. Likewise, preferential attachment algorithms are excellent models of network growth, branching and clustering, but these models are taciturn about the non-deterministic, dissipative character of natural processes, i.e., that evolution of a network affects its own course by molding the surrounding energy landscape.

Undoubtedly the physical portrayal of natural networks according to the statistical mechanics of open systems does not relate one-to-one with many mathematical models of networks. Specifically, when network nodes are depicted as identical with each other, there is no energetic bias that would structure the natural network. Likewise, when links are drawn as having equal capacities, there is no variation for natural selection to prefer a particular attachment. Also the notion that probability is physical may appear to some unmotivated and even bizarre, in particular since artificial neural networks adapt well to various input by adjusting weights. However, all forms of information embodied in physical presentations are also subjects of thermodynamics, hence the provided description is self-consistent when given only in energetic terms.

Finally, we would like to recall that our objective is neither to question established mathematical models and measures of networks nor conclusions founded on network theory, but to contribute to the discourse on complex systems by communicating the naturalistic tenet.

**Acknowledgments**

We thank Robert Leigh for valuable comments and corrections.